\documentclass[%
 pra,twocolumn,
superscriptaddress,
 amsmath,amssymb,
 aps,
floatfix,
]{revtex4-1}

\usepackage{graphicx}
\usepackage{dcolumn}
\usepackage{bm}
\usepackage[dvipsnames]{xcolor}
\usepackage{bbold}

\newcommand{\ndg}{{\phantom{\dagger}}}
\newcommand{\dg}{\dagger}
\newcommand{\ew}[1]{\left\langle #1 \right\rangle}
\newcommand{\ket}[1]{| #1 \rangle}
\newcommand{\bra}[1]{\langle #1 |}
\newcommand{\bracket}[2]{\langle#1|#2\rangle}
\newcommand{\ketbra}[2]{\left|#1\right\rangle\hspace{-1.1mm}\left\langle #2 \right|}

\newcommand {\hc}{\text{H.c.}}

\newcommand{\alex}[1]{ { #1 }}
\newcommand{\niki}[1]{ { #1 }}

\begin{document}

\preprint{APS/123-QED}

\title{\alex{Pronounced non-Markovian features in 
multiply-excited, multiple-emitter waveguide-QED: Retardation-induced anomalous population trapping}}

\author{Alexander Carmele}
\affiliation{%
Department of Physics, University of Auckland, Private Bag 92019, Auckland, New Zealand
}%
\author{Nikolett Nemet}
\affiliation{%
Department of Physics, University of Auckland, Private Bag 92019, Auckland, New Zealand
}%
\affiliation{%
Dodd-Walls Centre for Photonic and Quantum Technologies, New Zealand
}
\author{Victor Canela}
\affiliation{%
Department of Physics, University of Auckland, Private Bag 92019, Auckland, New Zealand
}%
\affiliation{%
Dodd-Walls Centre for Photonic and Quantum Technologies, New Zealand
}
\author{Scott Parkins}
\affiliation{%
Department of Physics, University of Auckland, Private Bag 92019, Auckland, New Zealand
}%
\affiliation{%
Dodd-Walls Centre for Photonic and Quantum Technologies, New Zealand
}
\date{\today}

\begin{abstract}
\alex{
The Markovian approximation is widely applied in the field of quantum optics due to the weak frequency dependence of the vacuum field amplitude, and in consequence non-Markovian effects are typically regarded to play a minor role in the optical electron-photon interaction.
Here, we give an example where non-Markovianity changes the qualitative behavior of a quantum optical system, rendering the Markovian approximation quantitatively and qualitatively insufficient. 
Namely, we study a multiple-emitter, multiple-excitation waveguide quantum-electrodynamic (waveguide-QED) system and include propagation time delay.}
In particular, we demonstrate anomalous population trapping as a result of the retardation in the excitation exchange between the waveguide and three initially excited emitters.
Allowing for local phases in the emitter-waveguide coupling, this population trapping cannot be recovered using a Markovian treatment, proving the essential role of non-Markovian dynamics in the process.
Furthermore, this time-delayed excitation exchange allows for a novel steady state, in which one emitter decays entirely 
to its ground state while the other two remain partially excited. 
\end{abstract}


\maketitle

\section{Introduction} 
One-dimensional (1D) waveguide-QED systems are attractive platforms 
for engineering light-matter interactions and
studying collective behavior in the ongoing efforts to construct scalable quantum networks 
\cite{booth2002counterpropagating,kimble2008the,petrosyan2008quantum,hafezi2012quantum,zheng2013persistent,zeeb2015superradiant,calajo2016atom,sollner2015deterministic,roy2017colloquium,pichler2017universal,lodahl2017chiral,bello2019unconventional}. 
Such systems are realized in photonic-like systems including  photonic crystal waveguides \cite{mekis1996high,colman2010temporal,goban2015superradiance,gonzalez2015subwavelength,aoki2009efficient,mahmoodian2017engineering,kim2018super-radiant}, optical fibers \cite{vetsch2010optical,goban2012demonstration,solano2017superradiance,white2019cavity,kato2019observation}, or metal and graphene plasmonic waveguides \cite{gonzalez2011entanglement,christensen2011graphene,bermudez2015coupling,mirhosseini2019cavity}.
Due to their one-dimensional structure, long-distance interactions become significant \cite{petrosyan2008quantum,zheng2013persistent,shahmoon2016highly}. 
As a result of these interactions mediated by left- and right-moving quantized electromagnetic fields, strongly entangled dynamics and collective, cooperative effects related to Dicke sub- and superradiance emerge
\cite{booth2002counterpropagating,serafini2006distributed,yudson2008multiphoton,aoki2009efficient,chang2012cavity,vanloo2013photon,gonzalez-ballestero2013non-markovian,zeeb2015superradiant,facchi2016bound,mirza2016two-photon,solano2017superradiance,song2018photon,white2019cavity,kato2019observation,bello2019unconventional}.
%

%
%
\begin{figure}[b!]  
\centering
\includegraphics[width=9cm]{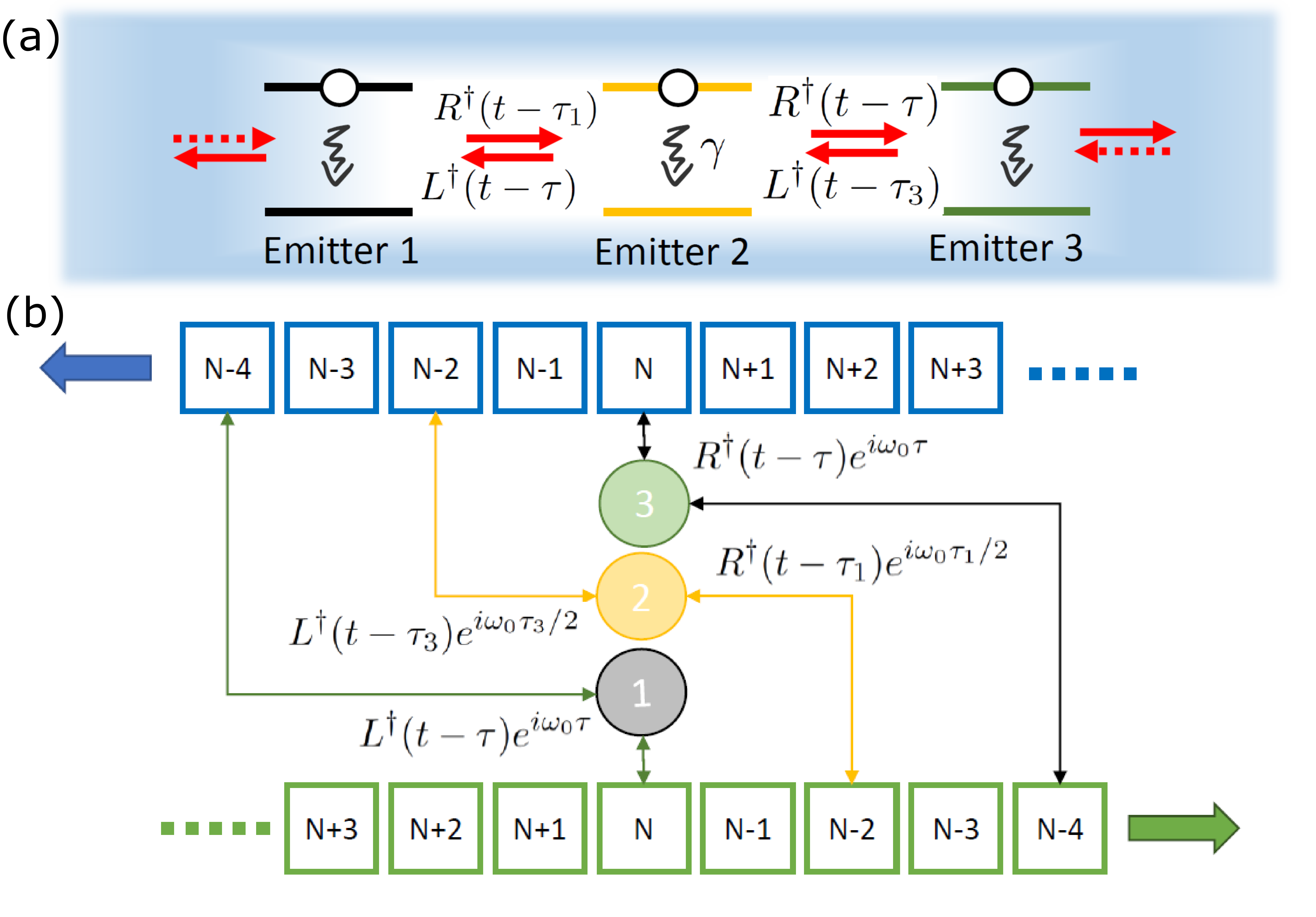}
\caption{Scheme of the simulated waveguide QED system. (a) The system consists of three identical emitter with transition frequency $\omega_0$ which couple to left- and right moving quantized light fields via the decay constant $\sqrt{\gamma}$. (b) Due to the delay (in the scheme two time steps $\tau_1=\tau_3=2 \Delta=2 \gamma^{-1}/10$ and $\tau=4\Delta$) a closed loop is formed between the first and third emitter interacting with their respective past bins. The interaction strongly depends on the phases $\omega_0\tau_1$ and $\omega_0\tau_3$.}
\label{fig:scheme}
\end{figure} 

%
\alex{In the framework of standard quantum optics, these systems are widely explored in the Markovian, single-emitter or single-excitation limit} \cite{PhysRevLett.123.123604,yudson2008multiphoton,roy2017colloquium,stolyarov2019few,dinc2018exact,denning2019quantum}. 
Such limits can be described by a variety of theoretical methods including real-space approaches \cite{PhysRevLett.123.123604,zheng2013persistent,fang2015waveguide,calajo2019exciting}, a Green's function approach \cite{hughes2017anisotropy,liao2016photon,see2017diagrammatic,schneider2016green}, Lindblad master equations \cite{shi2015multiphoton,baragiola2012n}, input-output theory \cite{xu2015input,fan2010input,fischer2018scattering,caneva2015quantum,witthaut2010photon}, and the Lippmann-Schwinger equation \cite{zheng2010waveguide,zheng2012strongly,shen2007strongly}.
Already in these regimes, exciting features 
have been predicted.
For example, strong photon-photon interactions 
can in principle be engineered, allowing for quantum computation protocols using flying qubits (propagating photons) and multilevel atoms \cite{zheng2013persistent,zheng2012strongly,paulisch2016universal,pichler2016photonic,lodahl2017chiral}.
Furthermore, bound states in the continuum are addressed via a joint two-photon pulse, showing that excitation trapping via multiple-photon scattering can occur without band-edge effects or cavities \cite{zheng2010waveguide,calajo2016atom,shi2016bound,calajo2019exciting}.
\alex{Beyond the single-excitation and/or single-emitter limit, the Markovian approximation becomes questionable and the aforementioned methods problematic \cite{gardiner2004quantum,breuer2002theory,breuer2016colloquium,devega_review,carmele_review,wignerdelay,samir,yumian}.}
In this work, \alex{we employ the matrix-product state representation to study exactly this regime, the multiple-excitation and  multiple-emitter limit.}
We focus, in particular, on the three-emitter and three-photon case, treating the emitters as two-level systems, which couple to the left- and right-moving photons and thereby interact with each other, subject to time delays associated with the propagation time of photons between emitters \cite{sinha2019non,paulisch2016universal,ramos2016non}. 
We choose throughout the paper the triply-excited state as the initial state and compare the relaxation dynamics in the Markovian and non-Markovian cases.
To compare both scenarios on the same footing, we employ the quantum stochastic Schr\"odinger equation approach \cite{gardiner2004quantum,gardiner1992wave,carmichael2009statistical} and numerically solve the model using a matrix-product-state algorithm \cite{guimond_dimerization,pichler2016photonic,lu2017intensified,droenner2019quantum,PhysRevLett.123.013601,schon2007sequential} as an alternative to the t-DMRG method in position space \cite{devega}.
We report on striking differences between the Markovian and non-Markovian description.
First, we find that in the case of non-Markovian excitation exchange, the triply-excited initial state allows for population trapping, in strong contrast to the Markovian description.
Second, time-delayed excitation exchange allows for anomalous population trapping, in which one emitter relaxes completely into its ground state while the two other emitters form a singly-excited dark state together with the waveguide field in between.
No local phase combination in the Markovian case allows for such anomalous population trapping, rendering the non-Markovian description qualitatively and quantitatively different from a Markovian treatment.

\section{Model} 
%
To demonstrate the importance of retardation-induced effects
and the underlying non-Markovian dynamics, we choose a system
consisting of three identical emitters with transition frequency $\omega_0$.
All three emitters interact with left- ($l^{(\dagger)}_\omega$) and right-moving photons ($r^{(\dg)}_\omega$) in a one-dimensional waveguide, as depicted in Fig.~\ref{fig:scheme}(a).
To focus on the retardation-induced effects, we neglect out-of-plane losses which inevitably enforce a fully thermalized, trivial steady-state in the ground state, and render the non-Markovian effects a transient, nevertheless important feature for waveguide-based counting experiments.
The Hamiltonian governing the free evolution of the combined, one-dimensional waveguide photon-emitter system reads:
\begin{align}
H_0/\hbar &= 
\omega_0
\sum_{i=1}^3  \sigma^{22}_i
+
\int d\omega \ \omega 
\left( 
r^\dg_\omega r^\ndg_\omega + l^\dg_\omega l^\ndg_\omega 
\right) ,
\end{align}
where the emitters are treated as two-level systems, with $\ket{1}$ as the ground state and $\ket{2}$ as the excited state, and with $\sigma^{ij}_n:=\ket{i}_{n n} \bra{j}$, the flip operator of the n-th emitter.
The interaction Hamiltonian 
describes the emitters interacting with right and left moving photons at the emitters' positions:
\begin{align}
H_I &= \notag
\hbar g_0
\sum_{i=1}^3 \sigma^{12}_i
\int d\omega \
\left( 
r^\dg_\omega e^{i\omega x_i/c}
+ l^\dg_\omega e^{-i\omega x_i/c} 
\right) + \hc, 
\end{align}
where we have assumed a frequency-independent coupling of the emitters to the quantized light field.
The position of the second emitter is chosen as $x_2=0$, leading to $x_1=-d_1/2=-c\tau_1/2$ for the first and $x_3=d_2/2=c\tau_3/2$ for the third emitter, with $c$ the speed of light in the waveguide.
After transforming into the interaction picture with respect to the free evolution Hamiltonian, and applying a time-independent phase shift to the left- and right-moving photonic field, the transformed Hamiltonian reads $H_I(t)=\hbar g_0\int  (H^{r,\omega}_I+H^{l,\omega}_I) d\omega$, where
\begin{align}\notag
H^{r,\omega}_I &= 
r^\dg_\omega(t)
\left( 
\sigma^{12}_1
+
\sigma^{12}_2
e^{-i\frac{\omega}{2}\tau_1}
+
\sigma^{12}_3
e^{-i\frac{\omega}{2}\tau}
\right)+ \hc, \\
H^{l,\omega}_I &= 
l^\dg_\omega(t)
\left( 
\sigma^{12}_1
e^{-i\frac{\omega}{2} \tau}
+
\sigma^{12}_2
e^{-i\frac{\omega}{2}\tau_3}
+
\sigma^{12}_3
\right)
+ \hc,
\end{align}
with $r^\dg_\omega(t)=r^\dg_\omega(0) \exp[i(\omega-\omega_0)t]$
and $l^\dg_\omega(t)=l^\dg_\omega(0) \exp[i(\omega-\omega_0)t]$,
and $\tau=(\tau_1+\tau_3)/2$, cf.~App.~\ref{app:hamiltonian}.
In the following, the left- and right-moving excitations are treated collectively: 
\begin{align}
R^\dg(t)=\int d\omega r^\dg_\omega(t), \quad
L^\dg(t)=\int d\omega l^\dg_\omega(t).
\end{align}
Given these definitions, the non-Markovian interaction Hamiltonian reads:
\begin{align}
\notag
H^\text{NM}_I(t)/\hbar &=
g_0
\left(
\sigma^{12}_1
\left( 
R^\dg(t)
+
e^{i\omega_0\tau}
L^\dg(t-\tau)
\right)
+
\hc \right) \\ 
&+
g_0
\left(
\sigma^{12}_2
R^\dg(t-\tau_1/2)
e^{i\frac{\omega_0}{2}\tau_1}
+
\hc \right)  \\ \notag
&+
g_0
\left(
\sigma^{12}_2
L^\dg(t-\tau_3/2)
e^{i\frac{\omega_0}{2}\tau_3}
+
\hc  
\right) 
 \\ \notag
&+
g_0
\left(
\sigma^{12}_3
\left( 
R^\dg(t-\tau)
e^{i\omega_0\tau}
+
L^\dg(t)
\right)
+
\hc
\right).
\end{align}
In the following, we compare the Markovian with the non-Markovian case.
The Markovian case neglects retardation effects between the excitation exchange, therefore in the Markovian approximation we set $R^{(\dg)}(t-t^\prime)\approx R^{(\dg)}(t)$ and $L^{(\dg)}(t-t^\prime)\approx L^{(\dg)}(t)$.
In this approximation, only the local phases but not the retardation in the amplitude are taken into account.
Consequently, the Markovian interaction Hamiltonian reads:
\begin{align}
\notag
&H^\text{M}_I(t)/\hbar = 
g_0
\left[
L^\dg(t)
\left(
\sigma^{12}_1
e^{i\omega_0\tau}
+
\sigma^{12}_2
e^{i\frac{\omega_0}{2}\tau_3}
+
\sigma^{12}_3
\right)
+ \hc
\right] \\
&+g_0
\left[
R^\dg(t)
\left(
\sigma^{12}_1
+
\sigma^{12}_2
e^{i\frac{\omega_0}{2}\tau_1}
+
e^{i\omega_0\tau}
\sigma^{12}_3
\right)
+ \hc
\right], 
\end{align}
where the emitters interact with time-local collective right- and left-moving fields and no time delay is present in the interaction.
We solve for the system's dynamics in both cases using the time-discrete Schr\"odinger equation with the time-step size $\Delta$ up to time $N\Delta$ in $N$ steps, cf.~Fig.~\ref{fig:scheme}(b):
\begin{align}
\label{eq:unitary_def}
\ket{\psi(n)}
&=U_\text{NM/M}(n,n-1)\ket{\psi(n-1)} \\ \notag
&=\exp\left[ 
-\frac{i}{\hbar}
\int_{(n-1)\Delta}^{n\Delta} H^\text{NM/M}_I(t^\prime) dt^\prime
\right]\ket{\psi(n-1)},
\end{align}
where $\Delta$ is small enough to minimize the error in the Suzuki-Trotter expansion
\cite{guimond_dimerization,pichler2016photonic,lu2017intensified,droenner2019quantum,PhysRevLett.123.013601,schon2007sequential}, and the evolution is taken either in the Markovian (M) or in the non-Markovian limit (NM).
\alex{
Here, the wavefunction is in MPS form:
\begin{align}\notag
\ket{\psi(n)}
=&\sum_{\substack{s,l_1\cdots l_N \\ r_1\cdots r_N  }}
 L^{[l_1]} R^{[r_1]} \cdots S^{[s]}
L^{[l_n]} R^{[r_n]} \mathbb{1}^{[l_{n+1}]}\mathbb{1}^{[r_{n+1}]} \dots \\ &\ket{l_1,r_1 \cdots s,l_n,r_n,l_{n+1},r_{n+1} \cdots l_N,r_N},
\label{eq:mps_representation}
\end{align}
where we assume vacuum input states for $m>n$ with identity tensor $L^{[l_m]}=\mathbb{1}^{[l_m]}$ and $R^{[r_m]}=\mathbb{1}^{[r_m]}$  in the time bins for the right- and left-moving field corresponding to a vacuum input state from left and right, and the indices $s$ count the degrees of freedom for the emitter system, and $r_i,l_i$ the number of excitations in the time bin for the left- and right-moving field.
In the following, we assume dim$[s]=2^3=8$, and choose a time-step size to guarantee that the dimension of the reservoir excitation does not exceed $\text{dim}[r_i]=\text{dim}[l_i]=3$. 
However, all results have also been calculated with $\text{dim}[r_i]=\text{dim}[l_i]=4$ to prove convergence.
}%
\begin{figure}[t!]  
\centering
\includegraphics[width=0.4\textwidth]{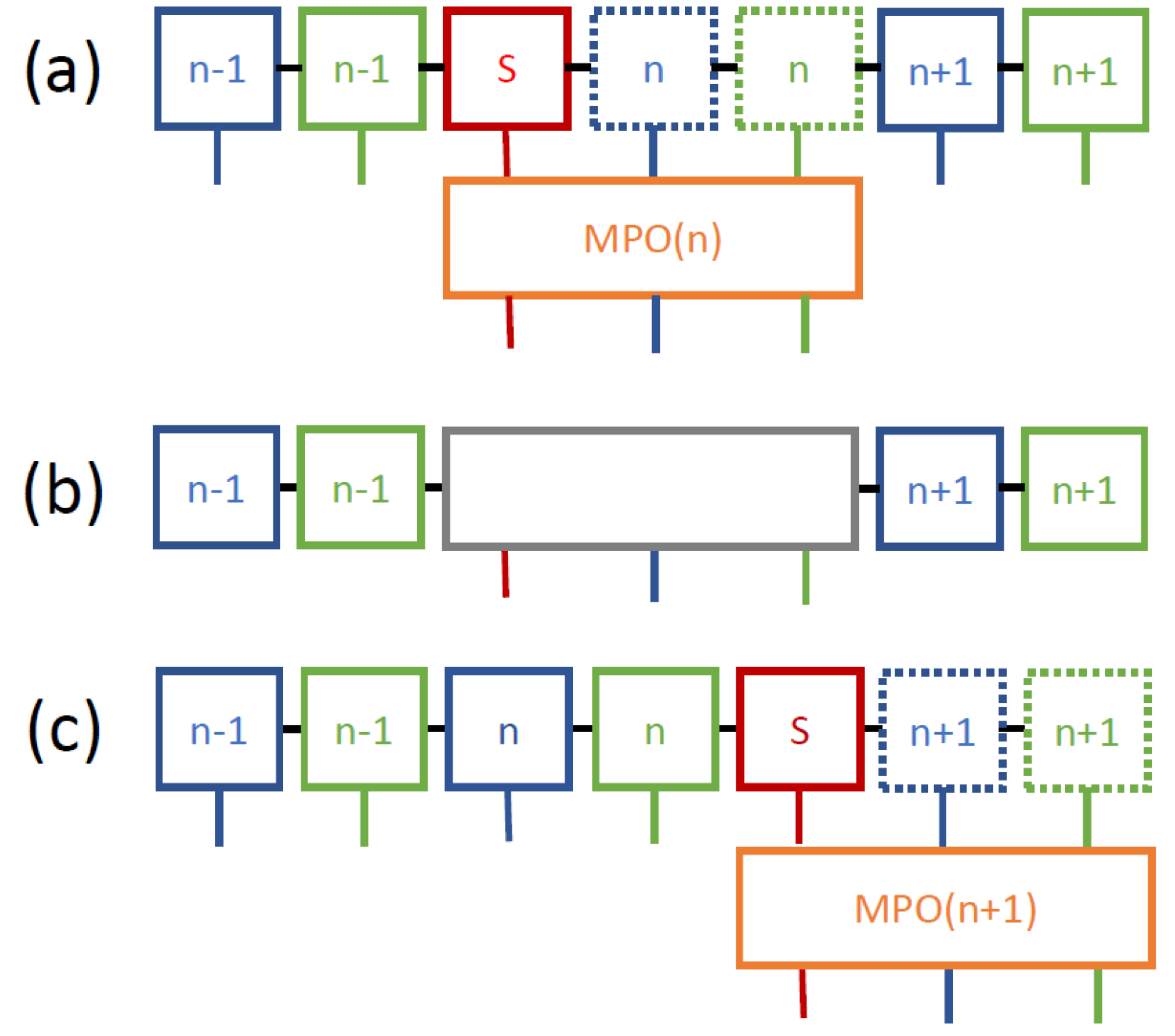}
\caption{Algorithm to compute the next time step via MPS in case of a Markovian dynamics, i.e. we set $R^{(\dg)}(n-m)\approx R^{(\dg)}(n)$ and $L^{(\dg)}(n-m)\approx L^{(\dg)}(n)$ and $m\Delta=\tau$.}
\label{fig:mpo_markovian}
\end{figure} 
To efficiently simulate the multiple-emitter, multiple-excitation case, we employ the matrix-product-state technique described in \cite{guimond_dimerization,pichler2016photonic,lu2017intensified,droenner2019quantum,PhysRevLett.123.013601,schon2007sequential}, and choose a collective basis for the flip operators of the emitters to allow for entangled initial states:\alex{$\ket{ijk} = \ket{(i-1)2^2 +(j-1)2^1+(k-1)2^0}$, which leads to, e.g., $\sigma^{12}_1 \equiv \ketbra{0}{4}+\ketbra{2}{6}+
\ketbra{1}{5}+\ketbra{3}{7}$, or $\ket{222}=\ket{7}$ the triply-excited state.}
We discuss in the corresponding sections the simulation protocol in detail.
%

\section{Markovian limit: No time delay}
We start our investigation in the Markovian limit, and calculate the system's dynamics with $U_\text{M}$ (Eq. \eqref{eq:unitary_def}) and the initial state $\ket{\psi(0)}=\ket{7}$ until the steady state is reached.
The Markovian case allows for a master-equation treatment with $g_0=\sqrt{2\pi\gamma}$ \cite{gardiner2004quantum,breuer2002theory,breuer2016colloquium,devega_review,carmele_review,wignerdelay,samir,yumian,kreinberg2018quantum}.
Tracing out the left- and right-moving photons leads to a  collective jump operator, 
$J:=\sqrt{\gamma}\left( \sigma^{12}_1 \exp[i\varphi_1] + \sigma_2^{12} + \sigma_3^{12} \exp[i\varphi_3] \right)$. 
The phases $\varphi_i$ can be chosen individually via local unitary transformations, or they arise from the spatial position without taking the finite distance into account in the evolution \cite{zeeb2015superradiant,ramos2016non}.
In the following, we nevertheless solve the dynamics 
using the quantum stochastic Schr\"odinger equation  
ignoring time-delay effects to 
give a Markovian evolution.
\alex{In Fig.~\ref{fig:mpo_markovian} the simulation protocol is depicted.
In the time-discrete basis, the time-local Hamiltonian in the matrix-product-operator MPO(n) form acts only on the present reservoir bins $n$ and the system state $s$ due to the Markovian approximation. 
After applying the evolution operator via contracting the physical indices (in the corresponding color code, red for the system bin, green for right moving field, blue for left moving field) in step (a), an entangled system-reservoir matrix is created.
To write the MPS in the canonical form, a Schmidt value decomposition is performed, and the entangled system-reservoir state is expressed as a matrix product (c), after which the next MPO(n+1) can be applied.
In this manner, the steady state can be calculated step by step in an efficient and less memory-consuming way.
Due to the Markovian approximation, i.e. we set $R^{(\dg)}(n-m)\approx R^{(\dg)}(n)$ and $L^{(\dg)}(n-m)\approx L^{(\dg)}(n)$ and $m\Delta=\tau$, no rearranging of the MPS is necessary.
The simulations are done, in the Markovian and non-Markovian case, until the steady state is reached.
In the MPS representation, (Eq.~\eqref{eq:mps_representation}), the steady state is reached when the application of the MPO leads to the identical tensor combination for every time step after the Schmidt value decomposition and swapping procedure, i.e. the MPO and subsequent re-arranging only acts as an index-shifting operator $l_n,r_n\rightarrow l_{n+1},r_{n+1}$.
}
\begin{figure}[t!]  
\centering
\includegraphics[width=0.5\textwidth]{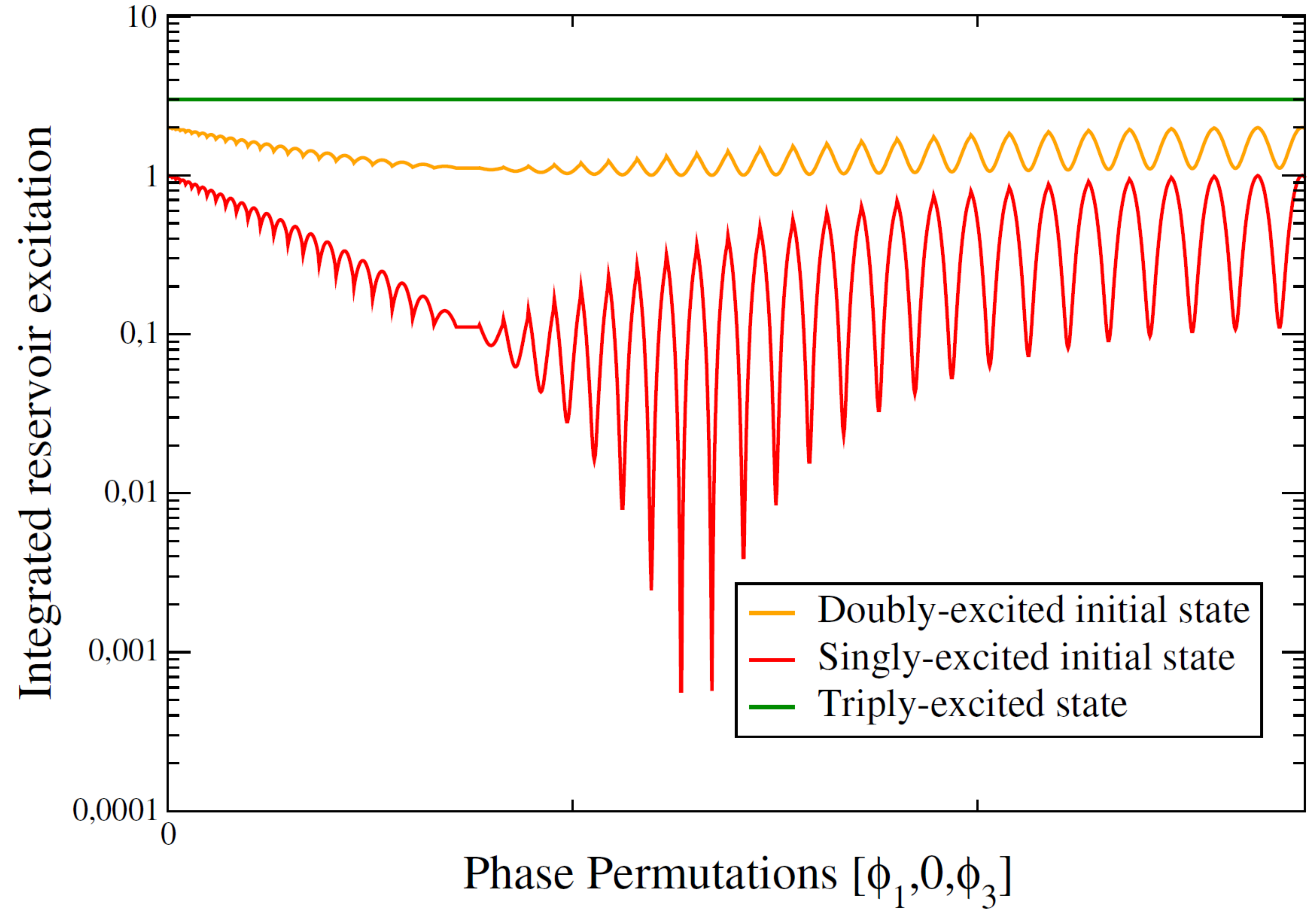}
\caption{The impact of different phase choices $\varphi:=(\varphi_1,\varphi_2,\varphi_3)=([0,2\pi],0,[0,2\pi])$ in the atom-waveguide couplings in the Markovian limit with photon operators: $R(t-t^\prime)=R(t)$ and $L(t-t^\prime)=L(t)$, on the integrated reservoir population in the steady state. If the system is initialized in the triply-excited state (green line), for all choices of phases, all excitation is radiated into the reservoir. If the system is initialized in a superposition of doubly-(orange line) or singly-excited states (red line), the only case where all excitation is radiated into the reservoir is when all phases are a multiple of $2\pi$.}
\label{fig:phase_permutation_3EX}
\end{figure} 
\begin{figure*}  
\includegraphics[width=\textwidth]{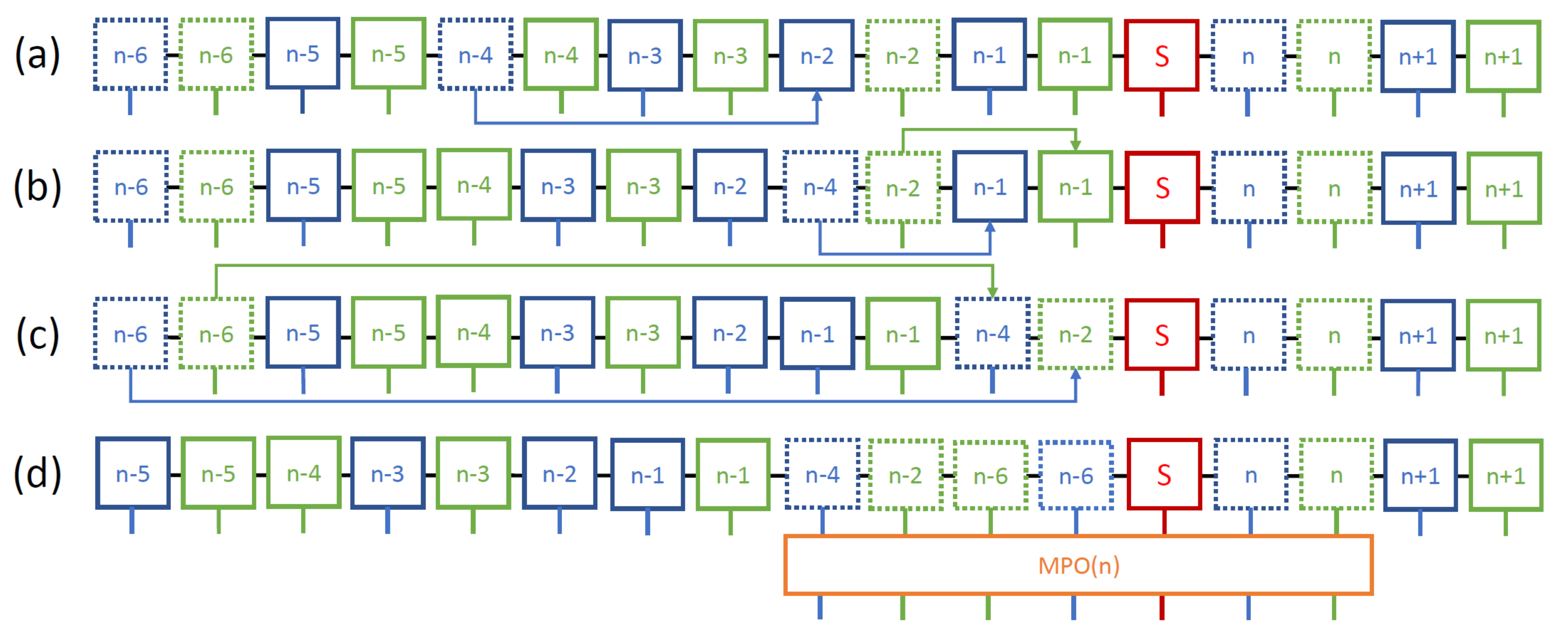}
\caption{Algorithm to compute the next time step via MPS in case of retardation of the left- (blue) and right-moving quantized field (green). The emitter $2$ (middle) interacts with a time delay of $2\Delta$ with the right-moving field, taking emitter $1$ as point of reference for the right-moving field, and emitter $2$ also interacts with the emitted left-moving field of emitter $3$, taking emitter $3$ as reference point for the left-moving field, resulting in a delay of $4\Delta$. The dashed squares couple to the system $s$ at time $n\Delta$. Since the canonical form of the MPS needs to be maintained, swapping procedures need to be applied where the orthogonality center is swapped from $n-6$ next to the system.}
\label{fig:mpo_non_markovian}
\end{figure*} 
In Fig.~\ref{fig:phase_permutation_3EX}, the phase dependence of the integrated reservoir excitation $I$ in the steady state is plotted for different initial states: $I=\sum_{n=0}^{N-f} \ew{R^\dg(n)R(n)} + \ew{L^\dg(n)L(n)}$ with $N$ as the number of time steps to reach the steady state and $f=(\tau_1+\tau_3)/\Delta$.
\alex{The phases are permutated by changing $\varphi_1,\varphi_3$ between $0$ to $2\pi$.}
If the system is initialized in the triply-excited state $\ket{\psi(0)}=\ket{7}$, the phases have no impact at all on the steady-state values and all excitation will eventually be radiated into the reservoirs on the left 
of emitter one and on the right 
of emitter three, leading for all phase permutations to the integrated reservoir occupation of $3$, cf. Fig.~\ref{fig:phase_permutation_3EX} (green line).
In contrast to the triply-excited case, the steady states of the emitters initially in a superposition of singly- ($\ket{1}+\ket{2}+\ket{4}$, red line) and doubly-excited states ($\ket{3}+\ket{5}+\ket{6}$, orange line) are strongly influenced by the choice of phases.
For those initial states, only if the phase difference vanishes, $\varphi_1=2\pi=\varphi_3$, is all radiation emitted into the reservoir. 
For all other phase combinations, population trapping occurs, and all emitters have a finite probability to be found in the excited state.
\alex{
Population trapping is created due to the fact that the individual decay of the emitters allows to populate a dark state of the Hamiltonian. 
In the two-emitter case, the standard light-matter Hamiltonian
can be written in the collective basis as:
\begin{align}
&H^s_{l-m} =
\hbar g_0 \sum_{i=1,2} ( \sigma^{12}_i + \sigma^{21}_i ) \\ \notag
&\equiv \hbar g_0 \left[ 
\ketbra{22}{21}+\ketbra{21}{11}+\ketbra{22}{12}+\ketbra{12}{11}+\text{h.c.}
\right]    
\end{align}
and therefore with $\ket{D}=(\ket{12}-\ket{21})/\sqrt{2}$ leads to such a dark state with $H^s_{l-m}\ket{D}=0$.
Therefore, with corresponding phase differences, such dark states can be driven via individual decays and lead to dark state population, or population trapping, e.g. \cite{petrosyan2008quantum,guimond_dimerization,hughes2017anisotropy,PhysRevA.100.023805,zeeb2015superradiant,white2019cavity,kato2019observation,carmele2014opto}.}
We conclude that, within the Markovian treatment, we find that either all emitters relax into their ground state or none.
For systems initialized in the triply-excited state, excitation trapping cannot be achieved.
And for the superradiant, symmetric singly-excited and doubly-excited initial states, the emitters 
undergo complete decay only in the case of vanishing phase difference.
We show now that including retardation and back-action effects changes this picture completely.
%

\section{Non-Markovian dynamics: Symmetric time delay.}
%
\alex{
In the non-Markovian case, the MPO is not only acting on the
present reservoir bins $n$ but also on the past left- and right-moving reservoir bins $n-m$ for $m>0$.
In Fig.~\ref{fig:mpo_non_markovian}, the simulation protocol is schematically explained for the case when emitter $2$ (middle) interacts with a time delay of $\tau_3/2=2\Delta$ with the emitted right-moving field of emitter $1$ and with a time delay of $\tau_1/2=4\Delta$ with the left-moving field from emitter $3$.
Correspondingly, emitter $1$ interacts with the present reservoir bin $n$ of the right-moving (green) and $n-6$ of the left-moving field (blue) to provide for a time delay $\tau=6\Delta$.
In Fig.~\ref{fig:mpo_non_markovian}, the dotted squares are those with which the MPO at time step $n$ interacts, and the orthogonality center is initially in $n-6$ of the left-moving field (blue).
For efficiency, the dotted squares need to be arranged in the MPS next to each other to avoid a memory-consuming contraction of the MPS without gain of information.
To ascertain the normalization, the swapping procedures start from left to right, cf. Fig.~\ref{fig:mpo_non_markovian}(a-b).
The swapping guarantees that the essential entanglement in between the reservoir bins is preserved.
Every swap creates a new orthogonality center, but in (c) the $n-6$ left-moving bin (blue) is swapped through the whole MPS next to the system $s$ and creates a well-defined orthogonality center before the MPO is applied.  
After applying the MPO, the resevoir bins are arranged in the previous, canonical form but the orthogonality center is left at $n-5$ (blue). 
In this protocol, the matrix product state remains in the canonical form and a fast numerical simulation is possible.}

We initialize the emitters in the triply-excited state and assume first symmetric delays, i.e. $\tau_1=\tau_3$.
In the case of quantum coherent feedback, the delay between the excitation exchanges introduces  a corresponding phase \cite{nemet2016enhanced,carmele2013single,kabuss2016unraveling,whalen2017open,julia,grimsmo2015time,guimond_dimerization,pichler2016photonic,lu2017intensified,droenner2019quantum}. 
In the following, we assume a transition frequency of the emitters to yield: $\omega_0\tau/2=2\pi$,
i.e. $\exp[\pm i\omega_0\tau]=1=\exp[\pm i\omega_0\tau/2]$.
We show now that the evolution under the influence of a finite delay in between the emission events together with subsequent back-actions from the previous emissions of each emitter lead to population trapping in strong contrast to the Markovian case.
\begin{figure}[t!]  
\centering
\includegraphics[width=.48\textwidth]{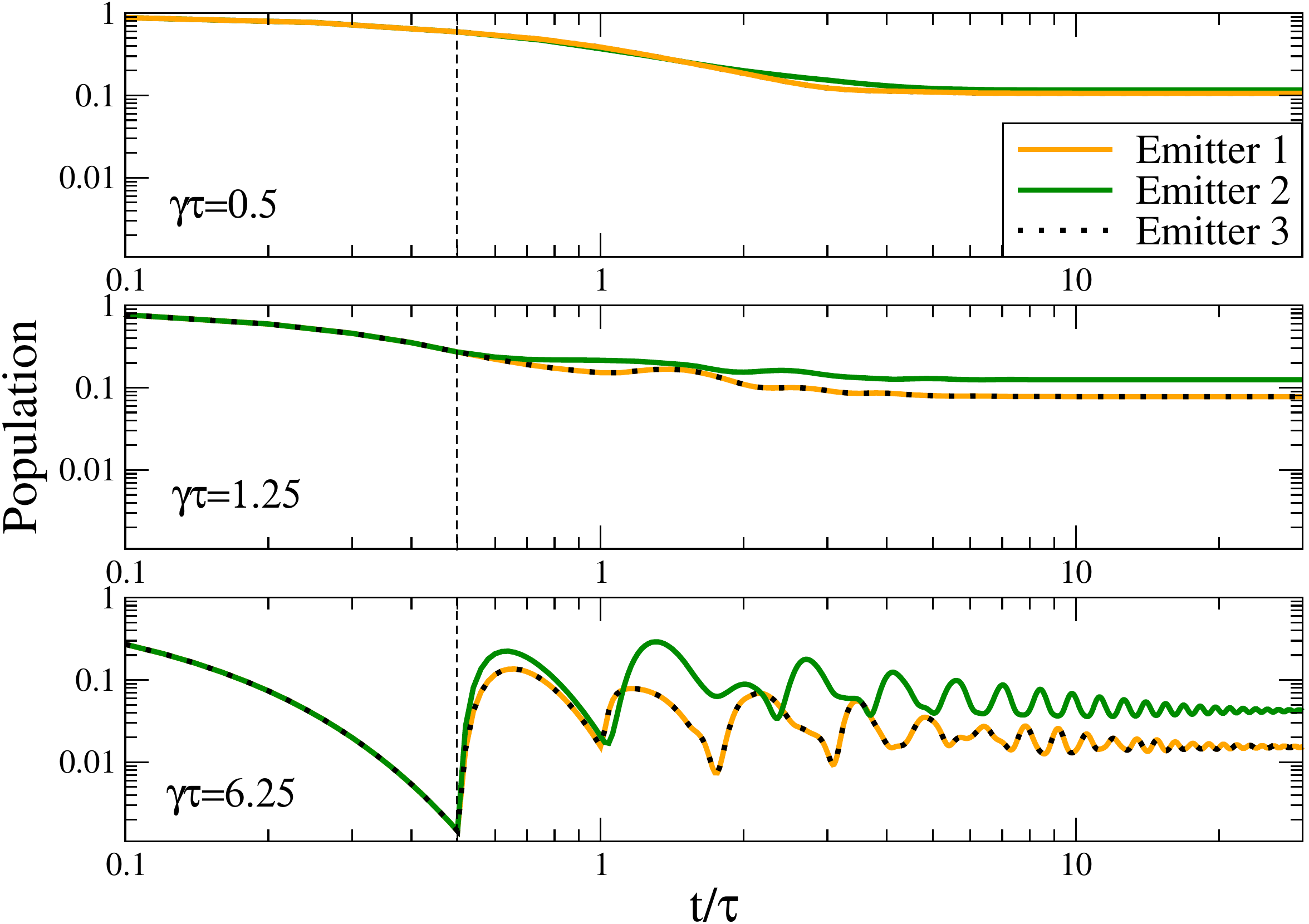}
\caption{The dynamics of the emitter populations for different feedback lengths with phase $\omega_\niki{0}\tau/2=2\pi$ for a system initially in the triply-excited state. Already short feedback times ($\tau=2$ns, i.e. $\gamma\tau=0.5$) lead to population trapping in contrast to the Markovian case. For longer feedback ($\tau=25$ns, i.e. $\gamma\tau=6.25$), slowly decaying oscillations become visible.} 
\label{fig:init111_w_diff_fb}
\end{figure} 
In Fig.~\ref{fig:init111_w_diff_fb}, the emitter populations in the presence of coherent quantum feedback are shown, e.g., $\ew{\sigma^{22}_1}=\sum_{i=1}^4 |\bracket{2i-1}{\psi(t)}|^2$.
Even for short feedback in comparison to the decay time, i.e., $\gamma\tau=0.5$, population trapping is observed (upper panel).
Emitter one (black dotted) and three (orange solid line) exhibit the same dynamics due to the symmetry of the system.
Both start to decay exponentially as expected before the first excitation with a neighboring emitter takes place $t\in[0,\tau/2]$. 
From this moment on (indicated with a dashed line), the decay is slowed down considerably due to the re-excitation and re-emission dynamics.
For longer times, the emitter starts to partially interact with its own "past" and after several round trips the population in the emitter is stabilized and a dark state is formed for this particular chosen phase.
We emphasize that this anomalous population trapping depends on the presence of left- and right-moving photons and correspondingly to a feedback effect induced by those.
Emitter two (green line) has, for longer delays (middle and lower panel), a slightly higher population than emitters one and three due to excitations from both the left and right emitters (black and orange line).
For very long feedback, i.e., $\gamma\tau=6.25$, a regime where the feedback phase ceases to have a strong influence, we observe an interesting oscillatory 
behavior in the emitter populations due to the feedback and finite excitation, which 
settles eventually to a small but finite steady-state value.
\begin{figure}[t!]  
\centering
\includegraphics[width=.48\textwidth]{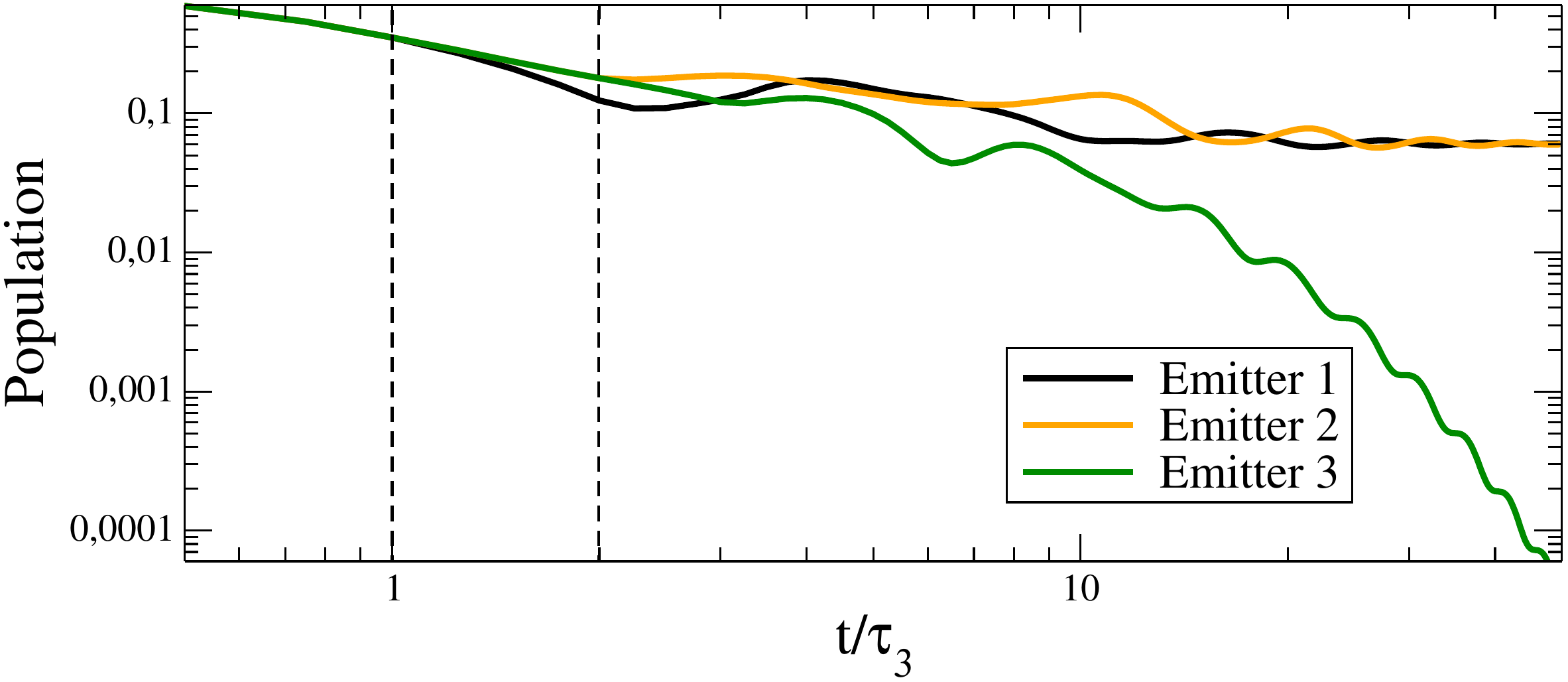}
\caption{The dynamics of the emitter populations if all emitters are initially in their excited state: $\ket{\psi(0)}=\ket{7}$ for a phase choice of $\omega_0(\tau_1+\tau_3)=3\pi$, and $\tau_1=2\tau_3$. Emitter 3 (green line) decays completely into its ground state while emitter 1 (black line) and 2 (orange line) remain in a partially excited state. This steady state is impossible to reach in the Markovian treatment if no additional interactions 
are included.}
\label{fig:init_superradiant_single_decay}
\end{figure} 
This example shows that allowing even for a short retardation and back-action time, the dynamics of the emitter populations changes qualitatively and quantitatively. 
Population trapping from an initial triply-excited emitter state cannot be recovered just with
local phases in the Hamiltonian.
This impossibility is lifted due to a time-delayed coherent 
feedback mechanism.
We emphasize that for very long delay the need for a particular choice of $\omega_0\tau/2=2\pi$ is partially lifted,
and it takes divergingly long for the emitter population to decay.
However, for long delay times $\gamma\tau\gg1$,
the population is also trapped in the reservoir between the emitters and the absolute stored population in the emitter system is exponentially small.
%

%
\section{Non-Markovian dynamics: Asymmetric time delay.}
Until now, we have discussed symmetric time delay $\tau_1=\tau_2$, or $|x_1|=|x_3|$.
Asymmetric time delay provides a further example how the full non-Markovian and quantized description of many-excitation dynamics in waveguide-QED deviates qualitatively from the Markovian treatment.
As shown above, in the Markovian treatment either all emitters remain partially excited or none of them do. 
For the triply-excited state, no population trapping occurs, and for other initial states the system is not able to reach a steady state 
with only one emitter in the ground state  
and the other emitters partially excited.
We show now that the non-Markovian description with asymmetric time delay allows for another example of anomalous population trapping, where one emitter decays completely into its ground state whereas the other emitters have a finite probability to be found in the excited state.

In Fig.~\ref{fig:init_superradiant_single_decay}, we choose a phase $\omega_0\tau=3\pi$ and delay times $\gamma\tau_1=1$ between the left (1) and middle emitter (2), and $\gamma\tau_3=0.5$ between the middle and the right emitter (3).
Excitingly, this setup allows for the right emitter (green line) to decay entirely to its ground state while the left (black line) and middle emitters (orange line) form a dark state together with the waveguide field in between and exhibit population trapping.
This effect results from the asymmetric delay between left- and right emission events.
For $t<\tau_3$, all emitters radiate unperturbed into the reservoir.
For $\tau_3<t<\tau_1$, the left emitter (black line) continues to radiate unperturbed whereas the middle and right emitters start to interact with the emitted photons.
Due to symmetry, both the right and middle emitters exhibit the same decay behavior for $t<\tau_1$.
This picture changes for larger times, as now the middle emitter's field starts to constructively interfere with the right-moving photons from emitter one. 
Emitter three interacts with its own past emission and decays faster, while emitters one and two start to form a superposition state. 
After several roundtrip times, $\gamma t\gtrsim 15$, emitter three has decayed, and no emission takes place. 
Interestingly, a necessary condition for this feature to happen is asymmetric feedback.
A symmetric feedback $\tau_1=\tau_3$ exhibits, as in the Markovian case, only finite population in all emitters, or none. 
This effect depends only on the destructive and constructive interference between left- and right-moving photons.
For different $\varphi=\omega_0\tau$, a different positioning needs to be chosen.
Quantity $\gamma\tau$ determines the extent of population trapping between emitter one and two, but not the qualitative effect. 
%

\section{Conclusion}
We have investigated a waveguide-QED system consisting of three emitters initialized in the triply-excited state, which interact via left- and right-moving photons.
We compared the Markovian and the non-Markovian case, i.e., without and with time delay in propagation between them. 
In the Markovian case, only a local phase is taken into account but no delayed amplitude in the re-emission events.
We recovered the well-known results, that the triply-excited state decays, independent of
phase choice, while the doubly- and singly-superradiant superposition state shows population trapping for any non-vanishing phase differences.
In strong contrast, a non-Markovian excitation exchange 
results in population trapping 
even if the system is initialized in the triply-excited state.
Furthermore, quantum feedback allows for states in which two emitters form a superposition state together with a part of the reservoir, whereas the third emitter relaxes entirely into the ground state; a state that is not possible to realize in a Markovian setup if only local phases in the jump operators, and no additional interactions, are assumed.
These examples prove the significance of time delay 
in many-emitter, many-excitation systems and the possibility of entirely new physics beyond the Markovian regime in the steady-state and 1D (or $\beta\rightarrow 1$) limit considered here, whereby the emitters radiate purely into the (detectable) waveguide modes, which is a regime already in reach of various waveguide-QED platforms \cite{beta1a,beta1b}.
\\ \ \\
AC gratefully acknowledges support from the Deutsche Forschungsgemeinschaft (DFG) through the project B1 of the SFB 910, and from the European Union’s Horizon	2020 research and innovation program under the	SONAR grant agreement no. [734690]. The calculations were performed using the ITensor Library \cite{ITensor}.

\begin{appendix}
\section{Hamiltonian of the multiple-emitter waveguide-QED system}\label{app:hamiltonian}
The free evolution Hamiltonian of the combined one-dimensional waveguide photons and emitters system reads:
\begin{align}
H_0/\hbar &= 
\sum_{i=\niki{1,2,3}} \omega_i \sigma^{22}_i
+
\int d\omega \ \omega 
\left( 
r^\dg_\omega r^\ndg_\omega + l^\dg_\omega l^\ndg_\omega 
\right) ,
\end{align}
where the emitters are treated as two-level systems with $\ket{1}$ as the ground state and $\ket{2}$ as the excited state, and $\sigma^{ij}_n=\ket{i}_{n n} \bra{j}$ for the n-th emitter, with $\sigma_n^\niki{12}=\ket{1}_{n n} \bra{2}$ the de-excitation operator of the n-th emitter.
The interaction Hamiltonian consists of the emitter interacting with right- and left-moving photons at the emitter's position:
\begin{align}
H_I/\hbar &= \notag
\sum_{i=1,2,3} \sigma^{12}_i
\int d\omega \
g_i(\omega)
\left( 
r^\dg_\omega e^{i\omega x_i/c}
+ l^\dg_\omega e^{-i\omega x_i/c} 
\right) \\
&+
\sum_{i=1,2,3} \sigma^{21}_i
\int d\omega \
g^*_i(\omega)
\left( 
r^\ndg_\omega e^{-i\omega x_i/c}
+ l^\ndg_\omega e^{i\omega x_i/c} 
\right).
\end{align}
The positions of the atoms are assumed to be centered around $x=0$, so in the case of three atoms, the middle atom is located at $x_\niki{0}=0$, the first (from left) atom at $x_\niki{1}=-d_1/2=-c\tau_1/2$, and the third atom at $x_\niki{3}=d_3/2=c\tau_3/2$.
Choosing a rotating frame corresponding to the middle atoms frequency $\omega_\niki{2}$ and the reservoir modes of the left and right-moving photons, the total Hamiltonian reads:
\begin{align}
H_I/\hbar &=
\delta_\niki{1} \niki{\sigma_1^{22}}
+
\delta_\niki{3} \niki{\sigma_3^{22}}
\\ \notag
&+
\sum_{i=1,2,3}
\left( 
\sigma^{12}_i
\int d\omega \
g^*_i(\omega)
r^\dg_\omega e^{i\omega \tau_i/2}
e^{-i(\omega_\niki{2}-\omega)t}
+ 
\hc \right) \\ \notag
&+
\sum_{i=1,2,3} 
\left(
\sigma^{12}_i
\int d\omega \
g^*_i(\omega)
l^\dg_\omega e^{-i\omega \tau_i/2} 
e^{-i(\omega_\niki{2}-\omega)t}
+
\hc
\right),
\end{align}
with $\delta_\niki{1}=\omega_\niki{1}-\omega_\niki{2}$ and
$\delta_\niki{3}=\omega_\niki{3}-\omega_\niki{2}$.
For convenience, we transform this Hamiltonian.
The left atom interacts without delay with
the right-moving field.
Also, we want the right atom to interact with
the left-moving field without delay:
To achieve this, we apply unitary transformations:
\begin{align*}
U_\niki{3} &=\exp\left[-\frac{i\tau_1}{2} 
\int d\omega \omega r^\dg_\omega r^\ndg_\omega
\right]
\rightarrow U_r r^\dg_\omega U_r^\dg=r^\dg_\omega e^{-i\niki{\omega}\tau_1/2} \\
U_\niki{1} &=\exp\left[-\frac{i\tau_3}{2} 
\int d\omega \omega l^\dg_\omega l^\ndg_\omega\right]
\rightarrow U_l l^\dg_\omega U_l^\dg=l^\dg_\omega e^{-i\omega\tau_3/2}.
\end{align*}%
Now, the interaction Hamiltonian reads:
\begin{align}
&H_I(t)/\hbar= \\ \notag
&=
\delta_\niki{1} \niki{\sigma_1^{22}}
+
\delta_\niki{3} \niki{\sigma_3^{22}}
\\
&+
\int d\omega \
\bigg(
\sigma^{12}_1
g^*_\niki{1}(\omega)
\left( 
r^\dg_\omega
+ 
l^\dg_\omega e^{-i\omega (\tau_1+\tau_3)/2} 
\right)
e^{-i(\omega_\niki{2}-\omega)t}
\\ \notag
&+
\sigma^{12}_\niki{2}
g^*_\niki{2}(\omega)
\left( 
r^\dg_\omega e^{-i\omega \tau_1/2}
+ 
l^\dg_\omega e^{-i\omega\tau_3/2} 
\right)
e^{-i(\omega_\niki{2}-\omega)t}
 \\ \notag
&+
\sigma^{12}_\niki{3}
g^*_\niki{3}(\omega)
\left( 
r^\dg_\omega e^{-i\omega (\tau_3+\tau_1)/2}
+ 
l^\dg_\omega 
\right)
e^{-i(\omega_\niki{2}-\omega)t}
+
\hc 
\bigg).
\end{align}
In the following, the left and right-moving
excitations are treated collectively,
and we assume the coupling elements to be constants with respect to frequency,
i.e., $g_i(\omega)=\sqrt{\gamma_i/(2\pi)}$. 
New operators are introduced: 
\begin{align}
R^\dg(t) &= \int d\omega r^\dg_\omega e^{i(\omega-\omega_\niki{2})t}/\sqrt{2\pi}, \\
L^\dg(t) &= \int d\omega l^\dg_\omega e^{i(\omega-\omega_\niki{2})t}/\sqrt{2\pi}.
\end{align}
Given these definitions, the interaction Hamiltonian reads:
\begin{align}
\notag
H_I(t)/\hbar &=
\delta_\niki{1} \niki{\sigma_1^{22}}
+
\delta_\niki{3} \niki{\sigma_3^{22}}
\\
&+
\sqrt{\gamma_\niki{1}}
\left(
\sigma\niki{^{12}_1}
\left( 
R^\dg(t)
+
e^{i\omega_\niki{2}\tau}
L^\dg(t-\tau)
\right)
+
\hc \right) \\ 
&+
\sqrt{\gamma_\niki{2}}
\left(
\sigma\niki{^{12}_2}
R^\dg(t-\tau_1/2)
e^{i\omega_\niki{2}\tau_1/2}
+
\hc \right)  \\ \notag
&+
\sqrt{\gamma_\niki{2}}
\left(
\sigma\niki{^{12}_2}
L^\dg(t-\tau_3/2)
e^{i\omega_\niki{2}\tau_3/2}
+
\hc  
\right) 
 \\ \notag
&+
\sqrt{\gamma_\niki{3}}
\left(
\sigma\niki{^{12}_3}
\left( 
R^\dg(t-\tau)
e^{i\omega_\niki{2}\tau}
+
L^\dg(t)
\right)
+
\hc
\right),
\end{align}
where $\tau=(\tau_1+\tau_3)/2$.
In the following, we use the notation for the collective states corresponding to $\ket{ijk} = \ket{(i-1) 2^2 +(j-1) 2^1+(k-1) 2^0 }$, which leads to, e.g., \niki{$\sigma^{12}_1=\ketbra{0}{4}+\ketbra{2}{6}+
\ketbra{1}{5}+\ketbra{3}{7}$}.
To solve the corresponding Schr\"odinger equation,
we switch to a time-discrete evolution picture, and
integrate from $0$ to $\Delta$ as the first time
step from $\ket{\psi(0)}$.
\begin{align}
\ket{\psi(n)} \notag
&=U(n,n-1)\ket{\psi(n-1)} \\ 
&=\exp\left[ 
-\frac{i}{\hbar}
\int_{(n-1)\Delta}^{n\Delta} H_I(t^\prime) dt^\prime
\right]\ket{\psi(n-1)}
\end{align}
We can now introduce in the discrete time-bin basis, the collective bath operators as:
\begin{align}
\Delta R^\dg(n) 
&= 
\int_{(n-1)\Delta}^{n\Delta}
\int d\omega r^\dg_\omega e^{i(\omega-\omega_\niki{2}) t}
\frac{dt}{\sqrt{2\pi\Delta}}
, \\
\Delta L^\dg(n) 
&= \int_{(n-1)\Delta}^{n\Delta}
\int d\omega l^\dg_\omega e^{i(\omega-\omega_\niki{2}) t}
\frac{dt}{\sqrt{2\pi\Delta}}.
\end{align}
Given the time bin dynamics, we can now study different cases.
Due to the richness of the model, we focus in the following only on the identical emitter case: $\gamma_i=\gamma$ and $\omega_\niki{1}=\omega_\niki{2}=\omega_\niki{3}\niki{=\omega_0}$. 
In this case, we have as free parameters only: $\gamma,\tau_1,\tau_3$, and the feedback phase $\omega_\niki{0} \Delta/2$ which determines all phases in the matrix-product-operator (MPO).
For example, if we set $\phi=\omega_\niki{0} (\tau_1+\tau_3)/2=2\pi$, in its feedback interaction the middle atom is subjected to the phase: $\phi_\niki{1}=\phi/(1+\tau_1/\tau_3)$ and $\phi_\niki{3}=\phi/(1+\tau_3/\tau_1)$.
For the symmetric case it follows obviously: $\phi_\niki{1,3}=\phi/2$.

\end{appendix}

\bibstyle{apsrev4-1}

\end{document}